# A 6.3pJ/b 30Mbps -30dB SIR-tolerant Broadband Interference-Robust Human Body Communication Transceiver using Time Domain Signal-Interference Separation

*Shovan Maity, Baibhab Chatterjee, Gregory Chang, Student Member, IEEE, Shreyas Sen, Senior Member, IEEE*
*School of Electrical and Computer Engineering, Purdue University*

*Abstract*— **Human Body Communication (HBC) provides a low power communication medium for energy constrained wearable/ implantable devices in and around the human body. This paper presents a broadband HBC transceiver implemented in 65nm CMOS that achieves 6.3pJ/b energy efficiency at 30Mbps with -30dB interference-tolerance. Capacitive termination at the receiver end is used to achieve a wideband HBC channel, and Time Domain Signal-Interference Separation (TD-SIS) using Integrating DDR (I-DDR) receiver allows a tolerance of -30 dB Signal to Interference Ratio (SIR) with a BER $<10^{-3}$. The transceiver achieves 18X improvement in energy-efficiency compared to the State-of-the-Art HBC transceivers while being simultaneously broadband (carrier-less, low-energy) and interference-robust. Such order-of-magnitude improvement in energy-efficiency and private communication through the human body may enable applications like closed-loop neuromodulation, health-monitoring, secure authentication among many others.**

## I. INTRODUCTION

Rapid advancement and miniaturization of semiconductor technology has enabled widespread availability of wearables, implantables, injectables, inhalables etc., which promise strong societal impact by enriching human lives. The small form factor of these devices makes them severely energy constrained due to their limited battery life. These energy-sparse devices are often interconnected using Wireless Body Area Network (WBAN), which often dominates the overall energy-budget. Human Body Communication (HBC), which utilizes the human body as a communication medium, has recently emerged [1]–[5] as a promising alternative to Wireless WBAN for interconnection of these implantable/wearable devices in and on the human body. Compared to the wireless medium, the human body provides a a) lower-loss b) broadband c) person-specific communication channel, enabling energy-efficient Broadband (BB) data transmission in HBC. This enables HBC to improve intra-body networking energy efficiency which enhances the lifetime of such devices or sometimes enable applications like closed-loop neuromodulation (electroceutical), health monitoring, human-computer interaction, secure authentication, augmented/virtual reality etc. High communication-energy is often a limiter for battery-size constrained BAN devices. For example high speed communication to an implantable using WBAN leads to fast battery drainage and surgeries for battery replacements more often than desired, which is one of the limiting factors in closed loop neuromodulation today. While promising as an energy-efficient BAN technique, previous HBC transceivers are either 1) Narrowband [1]–[3], [5] hence not completely utilizing the broadband HBC channel or 2) Broadband [4], but not simultaneously interference-robust at the highest data-rate. This paper presents a simultaneously BB and Interference-Robust

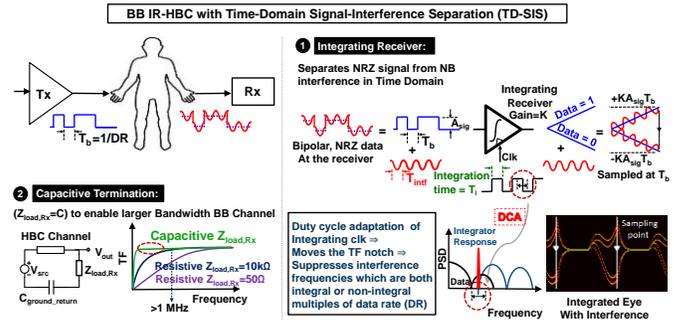

Figure 1: Broadband Interference-Robust (IR) HBC with TD-SIS. Two key techniques are used to enable BB HBC: (1) strong interference rejection using an Integrating Receiver with duty cycle adaptation (DCA) (2) larger bandwidth HBC channel using capacitive termination.

HBC (IR-HBC) Transceiver using Time-Domain Signal-Interference Separation (TD-SIS), achieving 6.3pJ/b 30Mbps -30dB SIR-tolerant operation, improving HBC energy-efficiency by over an order-of-magnitude compared to State of the Art transceivers. Such improvement in energy efficiency over traditional WBAN (>1nJ/b) and private communication (signal mostly contained within the body) can enable high speed communication in energy-constrained healthcare scenarios.

## II. BROADBAND INTERFERENCE ROBUST HBC

The HBC channel exhibits Broadband low-loss characteristics and broadband communication with 2-level NRZ data has been shown to be extremely energy-efficient [6] for wireline channels. However, one of the primary bottlenecks of implementing broadband HBC transceivers is due to the human body antenna effect, which results in the human body picking up environmental interference. The human body acts as an antenna in the tens-hundreds of MHz range and the ambient interference will close the NRZ eye for any broadband communication in this frequency range. State-of-the-art HBC transceivers primarily mitigate the interference problem through narrowband implementations by using static frequency bands with less interference [1] or by adaptively hopping between frequency bands by measuring channel quality [2]. However, such narrowband techniques require carrier and frequency up-down conversion similar to wireless systems, reducing energy-efficiency. As shown in Figure 1, Broadband Interference Robust HBC (IR-HBC) is achieved by using two key techniques: 1) Time Domain Signal Interference Separation (TD-SIS) through resettable integration operation and 2) Capacitive termination at the receiver end. Capacitive termination at the receiver creates a channel dominated by capacitive division of termination capacitance and ground return capacitance at low to moderate frequencies, allowing a higher bandwidth channel compared to resistive termination. This broadband channel enables carrier-less energy-efficient NRZ transmission. TD-SIS is implemented through a variable duty cycle Integrating Dual Data Rate (I-DDR) receiver for energy and area-efficient time domain separation of NRZ signal from narrowband (NB) interference. Intial work on IR-HBC using signal processing simulation was shown in [7] and adaptive duty cycle adaptation is shown in [8]. In this paper, we build on the initial theory, present a full tranceiver circuit and system that implements and augments to the theory, with IC design and measurement results *demonstrating the world's lowest-energy HBC transceiceiver reported in literature by over an order of magnitude.*

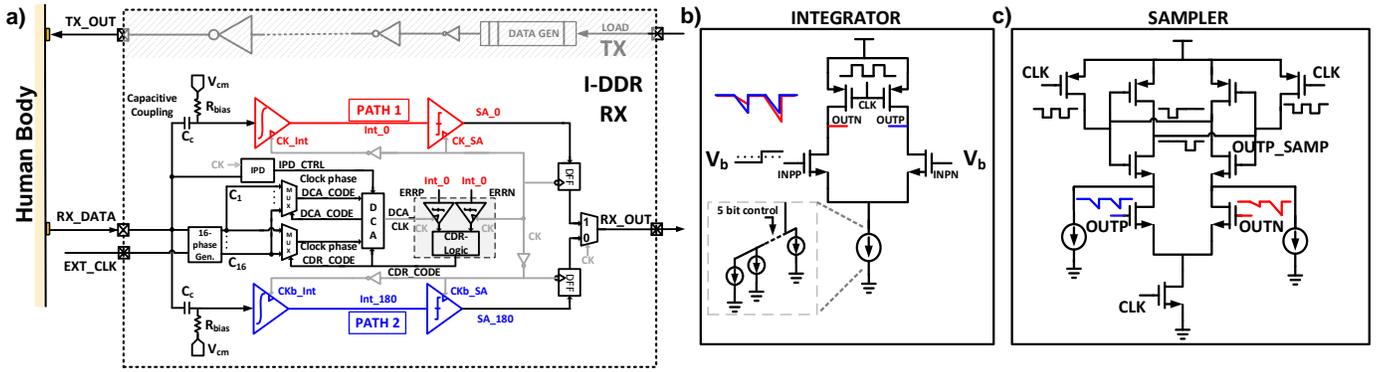

Figure 2: a) Block Diagram of the TD-SIS HBC Transceiver showing TX and I-DDR RX with Integrator, Sampler, Interference Period Detector (IPD), CDR and DCA b) Circuit level implementation of the resettable integrator and regenarative feedback sampler

*A. Theory of Time Domain Signal Interference Separation*

The interference affected received Broadband data ($S_{RX}$) can be represented as a combination of a NRZ data ($S_{sig}$) and a Narrow Band modulated sine wave ($S_{intf}$) as shown in equation (1).

$$S_{RX} = S_{sig} + S_{intf}$$

$$S_{sig}(t) = \pm A_{sig} \quad 0 \leq t \leq T_b$$
$$S_{intf}(t) = A_{intf}\sin(\omega_{intf}t + \varphi) \quad \forall t \quad (1)$$

A periodic integration with proper time-period ($T_i = nT_{intf} \leq T_b, n = integer$), serves as a transformation which separates the amplitude of these two signals in time domain, hence achieving TD-SIS, effectively creating a Sinc notch filter with sharp roll-off in frequency domain, tunable by varying the duty cycle of integration clock. For interferences that are integral multiple of data rate (DR), the integrated interference over a bit period is 0, whereas the integrated data is non-zero, opening the received eye as shown in equation (2), (3), (4).

$$IS_{RX}(T_b) = IS_{sig}(T_b) + IS_{intf}(T_b) \; (T_b = data\ period)$$

$$IS_{sig}(T_b) = \int_0^{T_b} S_{sig} = \pm K_{int}A_{sig}T_b \quad (2)$$

($K_{int} = integrator\ gain, A_{sig} = data\ amplitude$)

$$IS_{intf}(T_b) = \int_0^{T_b} S_{intf} = K_{int}\frac{A_{intf}\left[\cos(\varphi)-\cos\left(2\pi\frac{T_b}{T_{intf}}+\varphi\right)\right]}{\omega_i}$$
$$= 0, \; \forall T_b = nT_{intf}; n = positive\ integer \quad (3)$$

$$IS_{RX}(T_b) = IS_{sig}(T_b), \; \forall T_b = nT_{intf} \quad (4)$$

Interferences, which are a non-integral multiple of DR can be similarly suppressed by choosing the appropriate duty cycle (d) of the integration clock such that the integration time ($T_i<T_b$) is an integral multiple of $T_{intf}$.

$$IS_{intf}(T_i) = \int_0^{T_i} S_{intf} = K_{int}\frac{A_{intf}\left[\cos(\varphi)-\cos\left(2\pi\frac{T_i}{T_{intf}}+\varphi\right)\right]}{\omega_{intf}} \quad (5)$$

Now if $T_i$ corresponds to a duty cycle of $d = \frac{T_i}{T_{clk}} = \frac{T_i}{2T_b}$ and $T_b = kT_{intf}$, where $k$ can be a non-integer also then

$$IS_{intf}(T_i) = 0 \; \forall \; d = \frac{n}{2k} \; ; n = positive\ integer\ and\ d \leq \frac{1}{2}$$

So a variable duty cycle integrating receiver can be used to reject any narrowband interference signal by choosing proper duty cycle corresponding to a specific data rate of operation.

III. SYSTEM LEVEL DESIGN

The BB IR-HBC Transceiver (Figure 2a) consists of a NRZ transmitter, resettable I-DDR receiver, 16-phase clock generator, Clock Data Recovery (CDR) circuit with Error Samplers, Interference Period Detection (IPD) and Duty Cycle Adaptation (DCA) circuits. The transmitter can be configured to load external data, as well as generate PRBS output, and a series of inverters are used to drive the output. The two paths in the I-DDR receiver front end (Rx-FE) consist of a resettable integrator followed by a sampler each. If the reset operation is not performed between two successive integration operations then the integrated value at the end of a bit period will be affected by the integration of the previous bits, resulting in Inter Symbol Interference (ISI). So it is necessary to have reset after each integration operation, which in turn necessitates two separate paths to process consecutive bits. Each path works on opposite phases of the clock, enabling DDR operation. The integrated output must be sampled just before the end of the integration phase of the integrator. The final output is obtained by multiplexing the outputs of the individual paths. A baud rate Mueller Muller CDR takes input from two error detecting samplers and chooses the proper phase of a sixteen-phase clock generator. The IPD logic compares the period of the interference with the ON time of the integrating clock, and generates an adaptive control signal that adjusts the duty cycle through the DCA block to ensure $T_i = nT_{intf}$, maximizing the integrated eye-opening under strong interferences.

Figure 2b) and Figure 2c) shows the circuit level implementation of Integrator and Sampler. The integrator has a differential NMOS input stage and two PMOS switches act as the load. When the switches are ON, both the integrator outputs are connected to VDD and the integrator is in the RESET phase. During the EVALUATE phase, both the PMOS switches are turned OFF and the parasitic capacitance at the output nodes are discharged at a rate proportional to the input voltage of the NMOS stage. During an EVALUATE phase, since the parasitic capacitance is discharged through a constant current proportional to the input voltage, the differential output voltage is an integrated version of the differential input. The integrator gain is dependent on the $g_m$ of the input stage. This can be controlled by changing the transistor width of the input NMOS stage or by varying the tail bias current source value. However, the bias current value should be kept such that the NMOS input stage does not fall out of saturation during the integration period. So the bias current is kept adjustable through a 5b current DAC to maximize gain and prevent output saturation at any data rate. The differential integrator output is provided to a regenerative feedback sampler. The sampler has two back to back inverters connected in a chain. During the PRECHARGE phase the output of both the inverters are pulled high. During the evaluate phase the discharge rate of the inverter output is controlled by the sampler inputs. Due to the regenerative action one of the output will saturate to GND while the other will saturate to VDD. Offset generation capability is added to the sampler by steering current away from both outputs through 5 bit current DACs. Since the INTEGRATION phase of both the integrator and EVALUATE phase of the sampler are at the negative phase of the clock, they are supplied with out of phase clocks. Also the sampler evaluation needs to start before the finish of the integration period. To ensure there is a delay between the start of the EVALUATE phase of the sampler and the completion of the INTEGRATION phase of the integrator, the sampler clock is inverted and hence delayed and applied to the integrator. The DCA logic generates a duty cycled version of the integrating clock by performing an OR operation between the CDR clock phase and the DCA clock phase (generated using input from IPD). The generated duty cycled clock is directly applied to the sampler and the inverted clock is applied to the integrator. Since the integrating clock duty cycle needs to be reduced to achieve better interference rejection, the DCA generated clock duty cycle is required to be increased and OR logic is used to achieve that. The IPD logic compares ON times of clock and interference by time to voltage conversion through integration followed by sampling. The sinusoidal interference signal is converted to a square wave through a self-biased inverter based amplifier. Time to voltage conversion is achieved by using a constant current source to discharge a fixed capacitor through a switch, whose ON time is determined by either the clock signal or the amplified interference signal. During the OFF period the capacitor is charged back to VDD. An envelope detector is used to detect the peak level of the discharge from the two capacitors controlled by the clock and interference. The difference between these two values are compared through two comparators and if the difference is beyond a predetermined threshold then corrective action is taken by changing the DCA clock phase to change the duty cycle of the clock. The IPD output incorporates a dead-zone between the two thresholds to improve stability of the control loop.

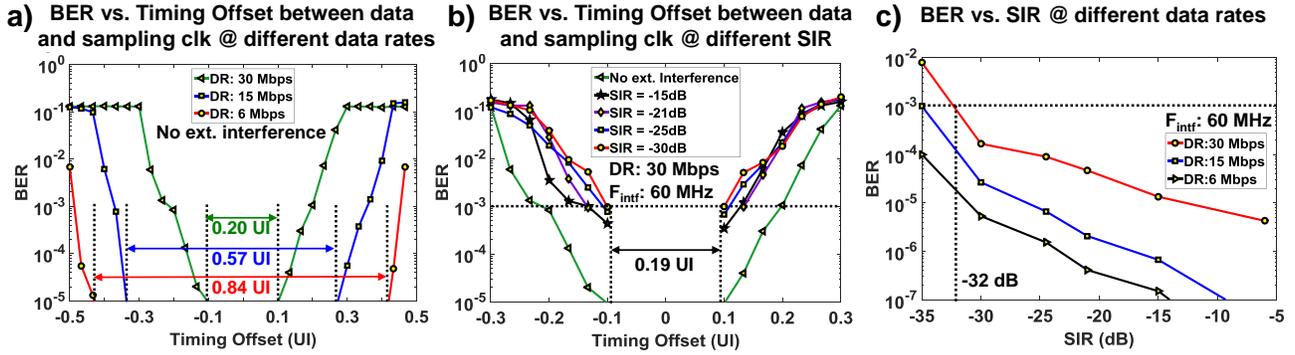

Figure 3: a) Measured BER bathtub plot at different data rates (DR) without any external interference b) BER bathtub plot at 30Mbps data rate under different Signal to Interference Ratios (SIR) c) BER performance at different signal to interference ratios (SIR) at different DR

## IV. MEASUREMENT RESULTS

The BB IR-HBC transceiver is fabricated in 65nm CMOS technology with 0.12mm$^2$ active area and the die is wire-bonded on a PCB for measurement. The receiver input capacitance is carefully minimized, as it affects HBC channel loss, and is dominated by required ESD capacitance. Bit-error-rate (BER) measurements is performed with PRBS data at multiple data rates and for multiple SIR at the highest data rate of

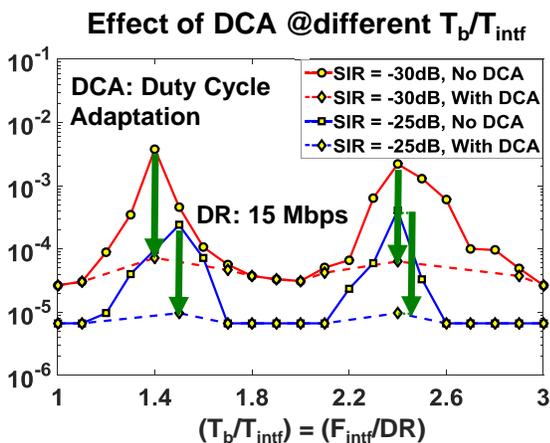

Figure 4: The effect of DCA on BER when the frequency of interference ($1/T_{intf}$) is not an integral multiple of the data rate ($1/T_b$) showing acceptable BER performance ($< 10^{-4}$) with any $T_{intf}$.

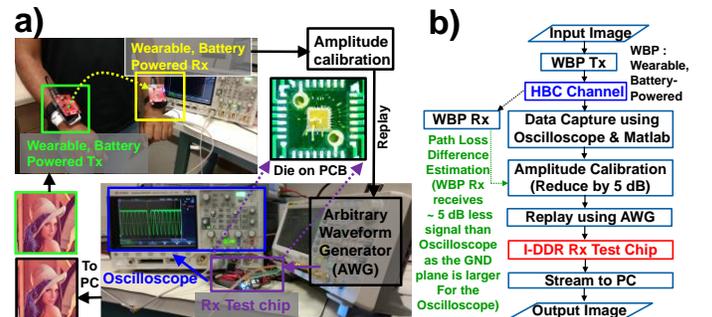

Figure 5: a) Setup of image transfer using HBC. b) Flowchart showing the stepwise experimental methodology. The AWG is used to replay the received signal after recalibrating the amplitude. The output from the chip is fed to a PC and the image is displayed through MATLAB. The PSNR of the received image is > 50dB.

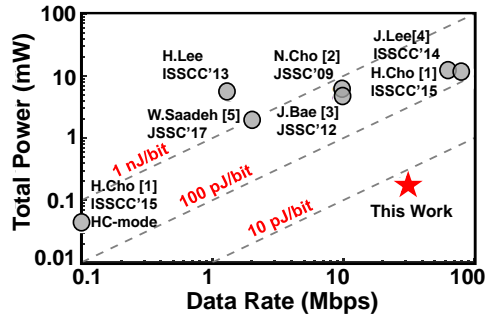

Figure 6: Comparison of total power (Tx+Rx) with state of the art HBC transceivers showing a measured energy efficiency < 10 pJ/bit.

30Mbps. Figure 3a shows the measured BER vs time offset (between data and clock) plot for different data rates without the presence of any external interference. It shows that there is a timing margin of 0.2UI (Unit Interval) for a BER of $10^{-5}$ at 30Mbps. Figure 3b shows the measured BER vs timing offset plot at the highest data rate of 30Mbps under different SIR conditions. There is a timing margin of 0.19UI for a BER of $10^{-3}$ even in the presence of -30dB SIR. The BER performance vs SIR plot for different data rate (Figure 3c) shows better interference tolerance at lower data rates. At 6Mbps data rate the achievable BER in presence of -35dB SIR is $10^{-4}$. Whereas the tolerable SIR for BER of $<10^{-3}$ across all data rates is -30dB.

Figure 4 shows the BER performance of the integrating receiver when the ratio of interference frequency and data rate ($T_b/T_{int}$) is varied from 1-3, under two different SIR conditions, and a fixed data rate of 15Mbps. It can be seen that at -30dB SIR the BER degrades to $10^{-2}$ for $T_b/T_{int}$ ratio of 1.5 and 2.5, without DCA. However with DCA turned on, there is >20X improvement in BER and the worst case BER is $<10^{-4}$, even with SIR=-30dB. Even in the absence of DCA, $<10^{-3}$ BER is achievable throughout the frequency range for SIR of -25dB, highlighting the robustness of IR-HBC.

Figure 5 demonstrates the experimental setup for transmission of an image and successful recovery of it through HBC. The signal is applied through an AWG by recalibrating the transmitted amplitude to reflect human body channel loss. The decoded output from the test chip is then captured in a PC and the image displayed through MATLAB. The received image has a PSNR of >50dB, corresponding to a BER of <$10^{-5}$. The energy-efficiency of the receiver improves with Data Rate with the best energy-efficiency of 3.27pJ/b at 30Mbps. Compared with state-of-the-art designs (Figure 6, Table I), BB IR-HBC achieves a transceiver energy-efficiency improvement of 18X (FoM=6.3pJ/b vs. 111.5pJ/b [1]), 4.4X smaller area than [5], best reported SIR-tolerance (-30dB at 30Mbps with BER<$10^{-3}$), while simultaneously achieving broadband and interference-robust operation without any external filters. The die micrograph along with the power consumption of different blocks and the active die area is shown in Figure 7.

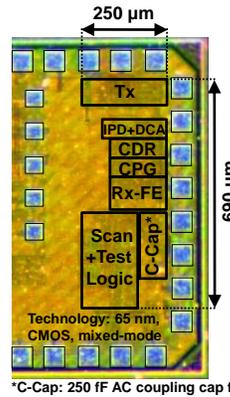

Figure 7: Die micrograph of the Tx and I-DDR Rx in 65 nm CMOS, and power and area consumption of individual components

## V. CONCLUSION

HBC is a promising alternative to wireless media for low power communication between connected devices around the human body. However interference due to human body antenna effect is one of the primary challenges of broadband HBC. This paper presents an interference robust broadband HBC transceiver in 65nm CMOS technology, which achieves energy efficiency of 6.3pJ/b at a data rate of 30Mbps and a SIR tolerance of -30dB for BER<$10^{-3}$. The achieved energy efficiency shows an 18X improvement over current state of the art HBC transceivers.

## VI. ACKNOWLEDGEMENTS

This work was supported in part by the National Science Foundation CRII Award under Grant CNS 1657455 and in part by the Air Force Office of Scientific Research YIP Award under Grant FA9550-17-1-0450.

## VII. REFERENCES

[1] H. Cho et al. "21.1 A 79pJ/b 80Mb/s full-duplex transceiver and a 42.5 uW 100kb/s super-regenerative transceiver for body channel communication," in *ISSCC 2015*

[2] N. Cho et al. "A 60 kb/s-10 Mb/s Adaptive Frequency Hopping Transceiver for Interference-Resilient Body Channel Communication," *IEEE J. Solid-State Circuits*, vol. 44, no. 3, pp. 708–717, Mar. 2009.

[3] J. Bae et al. "A 0.24-nJ/b Wireless Body-Area-Network Transceiver With Scalable Double-FSK Modulation," *IEEE JSSC*

[4] J. Lee et al. "30.7 A 60Mb/s wideband BCC transceiver with 150pJ/b RX and 31pJ/b TX for emerging wearable applications," ISSCC 2014,

[5] W. Saadeh et al. "A 1.1-mW Ground Effect-Resilient Body-Coupled Communication Transceiver With Pseudo OFDM for Head and Body Area Network," *IEEE J. Solid-State Circuits* 2017

[6] C. Thakkar et al. "23.2 A 32Gb/s bidirectional 4-channel 4pJ/b capacitively coupled link in 14nm CMOS for proximity communication," in *ISSCC 2016*

[7] S. Sen, "SocialHBC: Social Networking and Secure Authentication Using Interference-Robust Human Body Communication" *ISPLED 2016*

[8] S. Maity et al. "Adaptive interference rejection in Human Body Communication using variable duty cycle integrating DDR receiver" *DATE 2017*

|  | N.Cho [2] JSCC '09 | J.Bae [3] JSSC '12 | J. Lee [4] ISSCC '14 | H. Cho [1] ISSCC '15 | W. Saadeh [5] JSSC '17 | This Work |
|---|---|---|---|---|---|---|
| Process | 180nm CMOS | 180nm CMOS | 65nm CMOS | 65nm CMOS | 65nm CMOS | 65nm CMOS |
| Supply Voltage | 1 | 1 | 1.1 | 1.2 | 1.1 | 1 |
| Modulation | AFH FSK | Double FSK | 3-Level Walsh Coding | Coherent BPSK | 8 P-OFDM BPSK | NRZ |
| Maximum Data Rate | 10Mb/s | 10Mb/s | 60Mb/s | 80Mb/s | 2Mb/s | 30Mb/s |
| Tx Power | 2.4mW | 2mW | 1.85mW | 2.6mW | 0.87mW | 93uW |
| Rx Power | 3.7mW | 2.4mW | 9.02mW | 6.3mW | 1.1mW | 98uW |
| Energy/bit (Tx) | 240pJ/b | 200pJ/b | 31pJ/b | 32.5pJ/b | 435pJ/b | 3.1pJ/b |
| Energy/bit (Rx) | 370pJ/b | 240pJ/b | 150pJ/b | 79pJ/b | 550pJ/b | 3.27pJ/b |
| SIR (@ $10^{-3}$ BER) | -28dB | -20dB | -20dB | NA | NA | -32dB |
| Sensitivity | -65dBm @ $10^{-5}$ BER | -62dBm @ $10^{-5}$ BER | -58dBm @ <$10^{-5}$ BER | -58dBm | -83.1dBm @ $10^{-3}$ BER | -63.3*dBm @ $10^{-3}$ BER |
| Input Impedance | <100 Ω | 100-600 Ω | 10KΩ | _ | >> 50Ω | 22* KΩ, (Capacitive) |
| Area (mm²) | 2.3 | 12.5 | 1.12 | 5.93 | 0.542 | 0.122 |
| Interference Robust | Yes | Yes | Yes# | No | No | Yes |
| Broadband | No | No | Yes | No | No | Yes |

\# In robust mode using a separate band-stop filter, not at the highest data rate

\* Capacitive input termination, input impedance calculated at the Nyquist frequency of 15MHz corresponding to highest data rate of 30Mbps. Sensitivity corresponding to 6mV input swing for this input impedance

Table I: Performance Summary and comparison with related literature